\documentclass[aps,prl,twocolumn,amsfonts]{revtex4}

\usepackage{graphicx}

\usepackage{dcolumn}
\usepackage{amsmath}
\usepackage{psfig,epsfig}
\usepackage{times,mathptm}

\newcommand{\cm}{~cm$^{-1}$}

\newlength{\figw}
\begin{document}

\date{\today}

\title{ Simultaneous photon absorption as a probe of molecular
  interaction and hydrogen-bond correlation in liquids }

\author{Sander Woutersen}

\affiliation{Van 't Hoff Institute for Molecular Sciences, University
  of Amsterdam, Nieuwe Achtergracht~127-129, 1018~WV Amsterdam, The
  Netherlands}

\begin{abstract}
  We have investigated the simultaneous absorption of near-infrared
  photons by pairs of neighboring molecules in liquid methanol.
  Simultaneous absorption by two OH-stretching modes is found to occur
  at an energy higher than the sum of the two absorbing modes. This
  frequency shift arises from interaction between the modes, and its
  value has been used to determine the average coupling between
  neighboring methanol molecules. We find a rms coupling strength of
  $46\pm{}1$\cm, much larger than can be explained from
  transition-dipole coupling, suggesting that hydrogen-bond mediated
  interactions between neighboring molecules play an important role in
  liquid methanol. The most important aspect of simultaneous
  vibrational absorption is that it allows for a quantitative
  investigation of hydrogen-bond cooperativity. We derive the extent
  to which the hydrogen-bond strengths of neighboring molecules are
  correlated by comparing the line shape of the absorption band caused
  by simultaneous absorption with that of the fundamental transition.
  Surprisingly, neighboring hydrogen bonds in methanol are found to be
  strongly correlated, and from the data we obtain a hydrogen-bond
  correlation coefficient of $0.69\pm{}0.12$.
\end{abstract}

\maketitle

When two molecules are in close proximity, there is a finite
probability for them to simultaneously absorb a single photon and
'share' its energy. For vibrational transitions, this effect was first
observed more than fifty years ago in the near-infrared absorption
spectra of gas mixtures~\cite{fahrenfort54}, and later in crystalline
hydrogen chloride~\cite{ron63} and other condensed-phase
systems~\cite{bourderon,luck93}. Simultaneous absorption is observed
as an absorption peak at the sum frequency of the two vibrational
modes, and arises from a process in which one near-infrared photon is
absorbed, and both of the molecules involved become vibrationally
excited. It is obvious that simultaneous photon absorption can occur
only if the two molecules involved 'sense' each other's presence, that
is, if they interact~\cite{velsko80}. Here, we show that simultaneous
vibrational absorption in fact provides a unique probe of
intermolecular interactions in liquids.  It can be used to determine
the vibrational interactions between neighboring molecules, and more
importantly, it can be used to determine quantitatively the
correlation between the strengths of neighboring hydrogen bonds. This
hydrogen-bond correlation (or cooperativity) is believed to be a key
property of hydrogen-bonded
liquids~\cite{lamanna95,keutsch01,raiteri04}, but has as yet been
difficult to investigate experimentally.

We have investigated the simultaneous absorption by neighboring
CD$_3$OH molecules in liquid CD$_3$OH/CD$_3$OD mixtures.  By measuring
on isotopically diluted solutions of CD$_3$OH in CD$_3$OD, the
concentration of OH$\cdots$OH pairs can be varied continuously without
any change in the structural properties of the liquid. The absorption
spectra were measured using a Fourier-transform infrared spectrometer
(Bruker Vertex 70) with a spectral resolution of 0.5\cm.  Samples were
prepared by mixing appropriate amounts of CD$_3$OH and CD$_3$OD
(Euriso-Top~SA, $>$99.80\%D, HDO/D$_2$O content$<$0.03\%).  The weak
background absorption caused by overtones and combination modes of
CD-stretching vibrations~\cite{bourderon} was eliminated by
subtracting the spectrum of pure CD$_3$OD measured in the same sample
cell as used for the mixtures. Samples were kept between CaF$_2$
windows separated by either a 2.2 or 10.0~mm thick spacer.
\begin{figure}[h]
\centerline{\psfig{figure=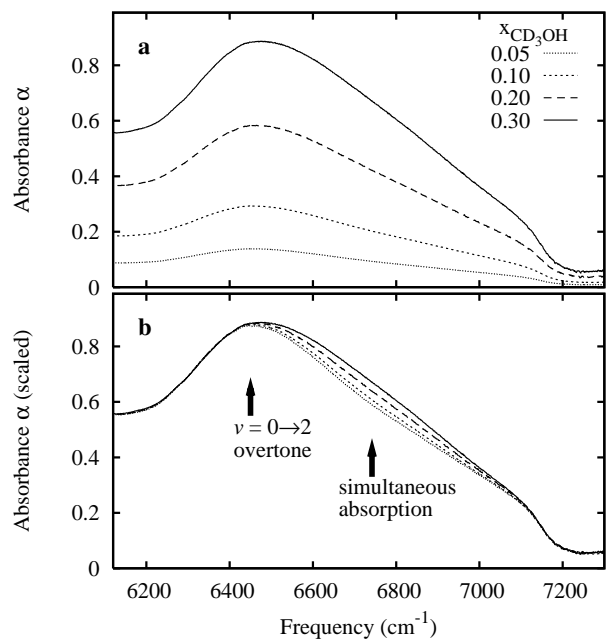,
    bbllx=-5,bblly=61,bburx=558,bbury=600,angle=270,
    width=\figw,clip=}}
\caption{(a)~Absorption spectra of CD$_3$OH/CD$_3$OD mixtures with
  increasing molar fractions $x_{\rm CD_3OH}$ (sample-cell thickness
  10.0~mm, CD$_3$OD absorption subtracted). (b) Same spectra, scaled
  to the OH-stretch overtone ($v=0\rightarrow2$) intensity at 6300\cm.
}
\label{fig:raw}
\end{figure}

Figure~\ref{fig:raw}a shows the near-infrared absorption spectrum of a
series of CD$_3$OH/CD$_3$OD mixtures with increasing molar
fraction $x_{\rm CD_3OH}$ (CD$_3$OD background
subtracted). The most prominent feature of these spectra is
the band centered at $\sim$6450\cm, which is caused by
$v=0\rightarrow2$ overtone absorption of the OH-stretch mode of
CD$_3$OH~\cite{bourderon}. The intensity of this band increases
linearly with the OH-stretch concentration $x_{\rm CD_3OH}$.  Closer
inspection shows that as the concentration of OH groups increases, a
second absorption feature appears at higher frequency. This
can most clearly be seen when the spectra are scaled to the overtone
absorption and overlayed, see Figure~\ref{fig:raw}b. As the
feature appears and grows with increasing CD$_3$OH concentration, it can
be assigned to simultaneous absorption by more than one CD$_3$OH
molecule~\cite{bourderon,luck93}.

\begin{figure}[h]
\centerline{\psfig{figure=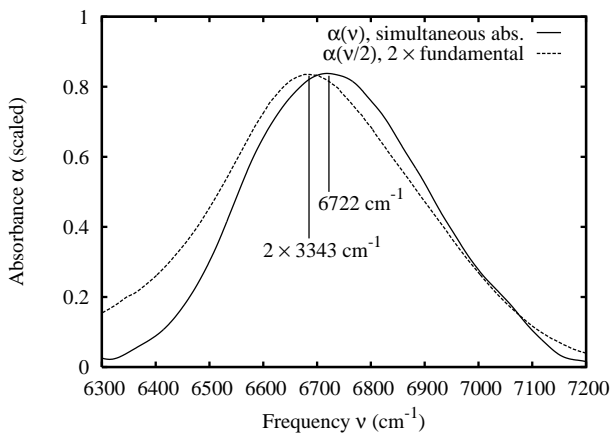,
    bbllx=70,bblly=65,bburx=559,bbury=757,
 angle=270, width=\figw,clip=}}
\caption{Solid curve: simultaneous absorption band,
  obtained by subtracting the absorption with $x_{\rm CD_3OH}=0.05$
  (scaled to overtone absorption) from that with $x_{\rm CD_3OH}=0.2$.
  Dashed curve: $v=0\rightarrow{}1$ absorption spectrum,
  multiplied horizontally by a factor of 2. The simultaneous
  absorption occurs at a frequency 36\cm\ higher than twice the
  fundamental frequency.}
\label{fig:shift}
\end{figure}

\begin{figure}[h]
\centerline{\psfig{figure=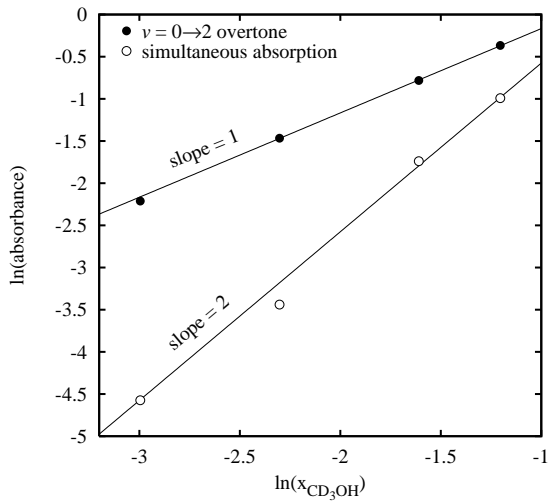,
    bbllx=67,bblly=61,bburx=560,bbury=600,
    angle=270,width=.9\figw,clip=}}
\caption{Log-log plot of the intensities of the $v=0\rightarrow{}2$ overtone
  band and of the simultaneous absorption band vs.\ $x_{\rm CD_3OH}$
  (the simultaneous absorption points
  have been vertically shifted by 1 unit for better comparison). Also
  shown are lines with slopes of 1 and 2, corresponding to linear and
  quadratic concentration dependence, respectively.  }
\label{fig:concdep}
\end{figure}

By subtracting the absorption spectrum at low CD$_3$OH concentration
(where the simultaneous absorption is relatively smaller) from the
spectra at higher $x_{\rm CD_3OH}$, the simultaneous absorption peak
can be observed separately from the more intense overtone band.  The
simultaneous absorption spectrum obtained in this way is shown in
Figure~\ref{fig:shift} (solid curve). Identical simultaneous
absorption spectra (apart from an overall scaling factor) are obtained
for all values of $x_{\rm CD_3OH}$ up to 0.3.
As the center frequency of the simultaneous absorption band
($\sim$6700\cm) is close to twice the fundamental OH-stretch frequency
of CD$_3$OH, it seems logical to assign the peak to simultaneous
absorption of one photon by the OH-stretch modes of two CD$_3$OH
molecules in close proximity~\cite{luck93}. To verify that
indeed two, and not more, CD$_3$OH molecules are involved, we have
determined the concentration dependence of the simultaneous
absorption (Figure~\ref{fig:concdep}). Whereas the overtone
absorption (solid points) varies linearly with $x_{\rm CD_3OH}$
confirming that it is due to
``one-molecule absorption'', the simultaneous absorption intensity
(open points) varies quadratically with $x_{\rm CD_3OH}$,
which proves that the simultaneous absorption involves
pairs of CD$_3$OH molecules (``two-molecule absorption''). It may be
noted that the quadratic $x_{\rm{CD_3OH}}$ dependence can already be
seen directly from the {\em linear} dependence of the simultaneous
absorption feature in the {\em scaled} data of Fig.~\ref{fig:raw}b. We
conclude that the observed simultaneous absorption peak must be due to
the absorption of photons by the OH-stretching modes of pairs of
CD$_3$OH molecules, since all other normal modes of CD$_3$OH have
frequencies too low to give a sum frequency of
$\sim$6700\cm~\cite{herzb}.

As already stated, simultaneous absorption can arise only if there is
interaction between the vibrational modes involved. In the present
case, the interaction required is between the OH groups of two
methanol molecules, which should therefore be in sufficiently close
proximity. It is easily demonstrated that only pairs of OH groups that
are direct neighbors can contribute significantly to the observed
simultaneous absorption band. The intensity of simultaneous absorption
is proportional to the square of the interaction between the
oscillators~\cite{velsko80}. As a consequence, the
simultaneous-absorption intensity due to dipolar interaction (the
dominant interaction mechanism~\cite{torii06b}) between two OH groups
decays with their distance as $r^{-6}$; and the non-dipolar
contributions to the interaction (occurring for instance through the
OH$\cdots$OH hydrogen bonds~\cite{belch83,woutnat}) decay even faster
with distance. Hence, the contribution of OH pairs that are
next-nearest neighbors (or even farther apart) is negligible, and the
simultaneous absorption is caused by neighboring OH groups only.

\begin{figure}[h]
\centerline{\epsfig{figure=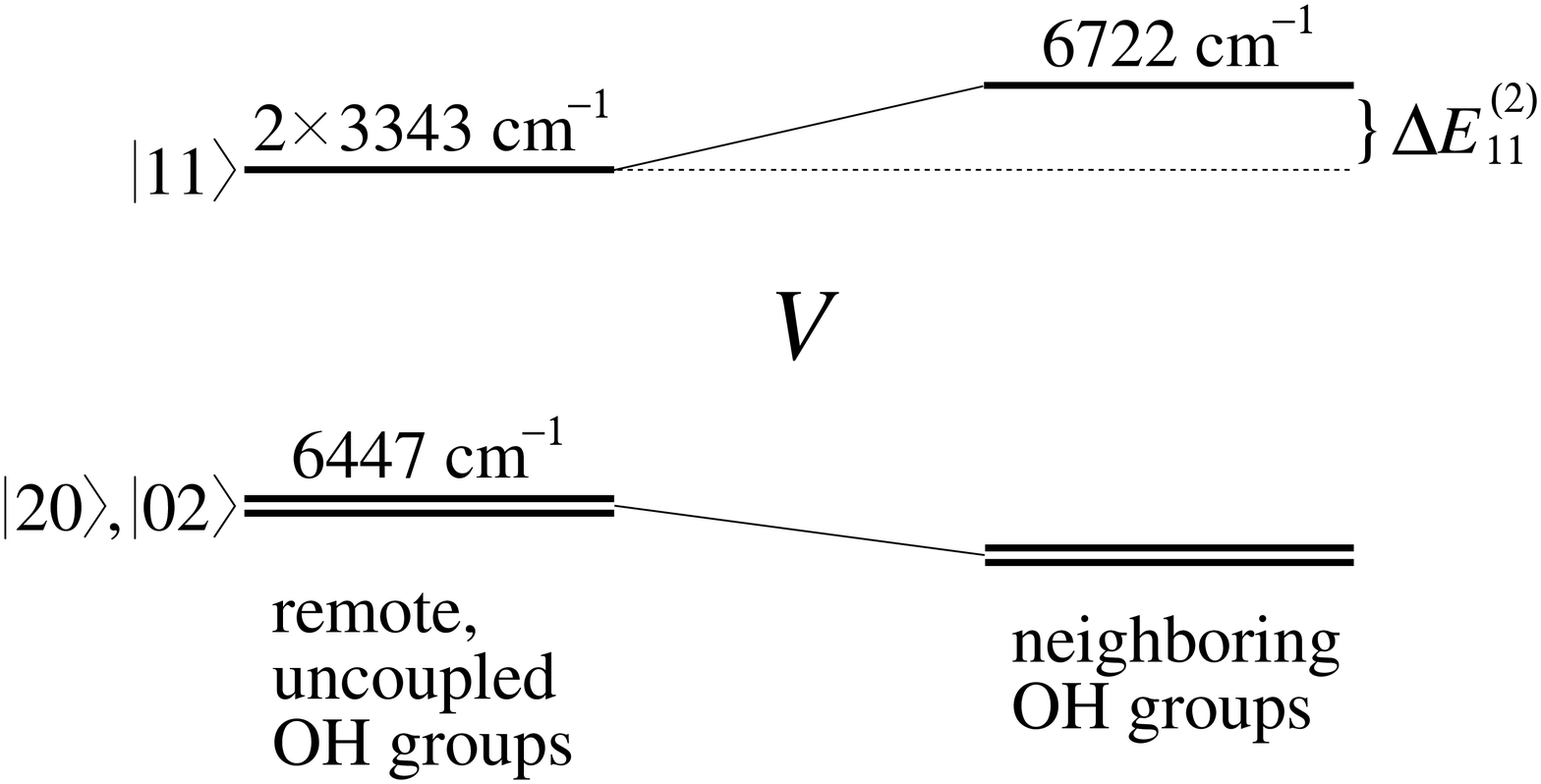,
    width=0.83\figw}}
\caption{Energy-level diagram (not to scale) for the OH-stretch
  two-exciton states. On the left the energy levels in absence of
  interaction between the two OH groups (isolated OH groups), on the
  right the energy levels when the molecules are neighbors.}
\label{fig:levels}
\end{figure}

Interestingly, the simultaneous absorption occurs not at exactly twice
the OH-stretch frequency ($2\times{}3343=6686$\cm), but at a frequency
which is higher (6722\cm). This can be seen clearly in
Figure~\ref{fig:shift}, where the simultaneous band is compared with
the fundamental absorption band multiplied horizontally by 2 (dashed
curve). This dashed curve corresponds to the hypothetical case of
simultaneous absorption by two non-interacting CD$_3$OH molecules
(note that because of the low $x_{\rm CD_3OH}$, the fundamental and
overtone bands are due to absorption by isolated CD$_3$OH molecules,
the concentration of which is much higher than that of paired ones).
Apparently, the interaction between the paired OH groups is large
enough to change the energy of the simultaneously excited state
significantly. From the position of the simultaneous absorption band
with respect to the doubled fundamental frequency (which corresponds
to the situation in absence of coupling), we can derive the (average)
magnitude of the coupling between the OH-stretching modes. This
coupling is dominated by the contribution lowest-order in the two
OH-stretch displacements~\cite{herzb}, and we can write it as $V =
\beta q_1 q_2,$ where $q_1$ and $q_2$ are the normal-coordinate
displacement operators of the neighboring OH-stretch modes, and
$\beta$ the coupling strength. We treat the coupling as a
perturbation, denoting by $|ij\rangle$ the unperturbed state having
the OH-stretch mode of molecule $A$ in the $v=i$ vibrational level,
and the OH-stretch mode of neighboring molecule $B$ in the $v=j$
vibrational level. Using harmonic approximations for the wave
functions~\cite{peter99,meth_noot1}, we have $\langle 11 | q_1 q_2 | 11
\rangle=0$, so that the first-order correction to the energy of the
simultaneously excited state $|11\rangle$ vanishes~\cite{meth_noot1}.
The second-order energy correction with respect to the situation
without coupling is given by~\cite{llquant}
\begin{equation}
\Delta E^{(2)}_{11}=\sum_{i,j}  \frac{
  \left|\langle ij|V|11\rangle\right|^2}
  {E^{(0)}_{11}-E^{(0)}_{ij}},
\end{equation}
where $E^{(0)}_{ij}$ is the unperturbed energy of state $|ij\rangle$.
Because of the $i\rightarrow{}i\pm{}1$ selection rule for the $q_i$
operator, the $V$ matrix element in the numerator is nonzero for only
a few states. Of these, the only ones sufficiently close in energy to
contribute significantly to the above sum are the $|02\rangle$ and
$|20\rangle$ overtone states, see Figure~\ref{fig:levels}. Hence, we
have
\begin{equation}
\left\langle \Delta E^{(2)}_{11}\right\rangle=\frac{
4\left\langle \beta^2 \right\rangle
 }{
E^{(0)}_{11}-E^{(0)}_{02}
},
\end{equation}
where we have used harmonic vibrational wave functions to calculate the
matrix element~\cite{llquant}, and where $\langle\ldots\rangle$
denotes ensemble averaging over all OH pairs contributing to the
simultaneous absorption band. Since only nearest neighbors contribute
significantly, this averaging is essentially over the relative
orientation distribution of two neighboring OH groups.
Setting $\langle\Delta E^{(2)}_{11}\rangle$ in the above expression
equal to the observed shift of $36\pm{}2$\cm, and substituting the unperturbed
values $E^{(0)}_{11}$ and $E^{(0)}_{02}$ (see Fig.~\ref{fig:levels}),
we obtain the root-mean-square
coupling constant $\langle\beta^2\rangle^{1/2}=46\pm{}1$\cm\cite{meth_noot2}.

This experimentally determined OH-OH interaction may be compared to
the transition-dipole coupling between the two OH groups, which is in
general the dominant contribution to intermolecular vibrational
couplings~\cite{moran04,torii06b}. The transition-dipole coupling
strength between two OH groups $A$ and $B$ is~\cite{moran04}
\begin{equation}
\beta_{\rm dipole}=\frac{1}{4\pi\epsilon_0}
\left[
    \frac{
        {{\pmb\mu}_A}\cdot{\pmb\mu_B}
    }{r_{AB}^3}
    -3\frac{
      ({\bf r}_{AB}\cdot{\pmb\mu}_{A})
            ({\bf r}_{AB}\cdot{{\pmb\mu}_{B}})
    }{r_{AB}^5}
\right],
\label{eq:dipole}
\end{equation}
where ${\bf{r}}_{AB}$ is the distance vector between the transition
dipoles ${\pmb\mu}_A$ and ${\pmb\mu}_B$. The values of
$\mu_A=\mu_B=0.264$~D and $r_{AB}=2.84$~\AA\ are
known~\cite{bertie97}, but the angle between the transition dipoles of
the two CD$_3$OH molecules will vary from pair to pair. To obtain an
upper limit for the transition-dipole coupling, we calculate
$\beta_{\rm dipole}$ for the case of parallel transition dipoles, for
which we find $\beta_{\rm dipole}= -30$\cm. This upper limit has a
magnitude much smaller than the observed rms value of 46\cm.
Hence, the coupling between two neighboring OH-groups is much larger
than can be explained from transition-dipole coupling alone, and
apparently other coupling mechanisms, like hydrogen-bond mediated
coupling~\cite{belch83,woutnat,moran04}, contribute strongly to the
molecular interaction in methanol.

\begin{figure}[h]
\centerline{\psfig{figure=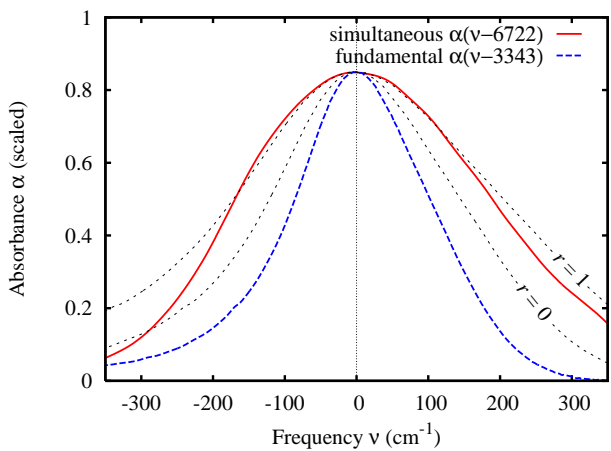,
 bbllx=70,bblly=65,bburx=559,bbury=732,
    angle=270,width=\figw,clip=}}
\caption{Solid curve: simultaneous absorption band. Dashed curve:
  fundamental absorption band. Both bands have been centered at the
  origin. Dotted curves: fundamental absorption band, horizontally
  multiplied by a factor of $\sqrt{2}$ and by a factor 2.  These two
  curves are the simultaneous absorption bands expected if the two
  coupled OH-oscillators are completely uncorrelated ($r=0$), and
  completely correlated ($r=1$), respectively. The experimental
  simultaneous absorption band is between the $r=0$ and $r=1$ curves,
  indicating that the OH-frequencies are partially correlated.}
\label{fig:corr}
\end{figure}

Perhaps the most important and surprising aspect of simultaneous
vibrational absorption is that it can be used to study correlation
among hydrogen bonds. There is strong
evidence~\cite{lamanna95,keutsch01,raiteri04} that in
hydrogen-bonded liquids, hydrogen bonding occurs in a cooperative way:
an OH group acting as an acceptor in a strong hydrogen bond in turn
tends to form strong hydrogen bonds as a donor, and {\it vice versa}.
It is well known that the OH-stretch frequency of a hydrogen-bonded OH
group is proportional with the strength of the hydrogen bond (low
OH-stretch frequency corresponding to short OH$\cdots$O hydrogen-bond
distance)~\cite{lawrence03b,fecko03}.  In hydrogen-bonded liquids,
there generally exists a distribution of hydrogen-bond strengths,
leading to an OH-stretch absorption band that is inhomogeneously
broadened~\cite{fecko03,cowan05}, its width reflecting the width of the
distribution of hydrogen-bond lengths. Because of the proportionality
between hydrogen-bond length and OH-stretch frequency, a correlation
between neighboring hydrogen bonds should give rise to an identical
correlation between the OH-stretch frequencies of neighboring OH
groups. Hence, by determining the correlation between the OH-stretch
frequencies $\nu_A$ and $\nu_B$ of two neighboring OH groups, we can
directly determine the correlation between their hydrogen bonds. To
what extent two frequencies $\nu_A$ and $\nu_B$ are correlated can be
quantitatively determined by comparing the width of the distribution
of the sum $\nu_A+\nu_B$ with that of the fundamental distributions of
$\nu_A$ and $\nu_B$~\cite{feller68}, that is, by comparing the widths
of the simultaneous and fundamental absorption bands. We define the
correlation coefficient between the frequencies $\nu_A$ and $\nu_B$ in
the usual manner as
\begin{equation}
  r=\frac
  {\left\langle \left(\nu_A-\left\langle\nu\right\rangle\right)
    \left(\nu_B-\left\langle\nu\right\rangle\right)\right\rangle}
  {\sigma_{\nu}^2},
\end{equation}
where $\langle\ldots\rangle$ denotes ensemble averaging, and where we have
defined $\langle\nu\rangle=\langle\nu_A\rangle=\langle\nu_B\rangle$
and $\sigma_{\nu}=\sigma_{\nu_A}=\sigma_{\nu_B}$ (note that this is
possible because the probability distributions of $\nu_A$ and $\nu_B$
are identical). It is easily shown that the correlation coefficient is
related to the ratio of the widths of the fundamental and simultaneous
absorption bands by~\cite{feller68}
\begin{equation}
  \frac {\sigma_{\nu_A+\nu_B}} {\sigma_\nu} = \sqrt{2+2r},
  \label{eq:ratio}
\end{equation}
a result that is independent of the functional form of the $\nu_{A,B}$
probability distribution. The familiar limiting cases of this equation
are: completely independent frequencies ($r=0$), where
$\sigma_{\nu_A+\nu_B}^2=\sigma_{\nu_A}^2+\sigma_{\nu_B}^2=2\sigma_\nu^2$
so that $\sigma_{\nu_A+\nu_B}=\sqrt{2}\sigma_\nu$, and completely
correlated frequencies ($r=1$), where $\nu_A=\nu_B$ so that
$\sigma_{\nu_A+\nu_B}=\sigma_{2\nu_A}=2\sigma_{\nu}$. The simultaneous
absorption bands predicted for these two limiting $r$ values are drawn
as dotted curves in Figure~\ref{fig:corr}. The observed simultaneous
absorption band lies between these limiting cases, which implies that
the OH-stretching frequencies of neighboring methanol molecules are
partially, but not completely correlated. From the observed ratio
$\sigma_{\nu_A+\nu_B}/\sigma_{\nu}$ of the widths of the simultaneous
and fundamental absorption bands we can determine $r$ by means of
Eq.~(\ref{eq:ratio}). We observe
$\sigma_{\nu_A+\nu_B}/\sigma_{\nu}=1.84\pm{}0.06$, from which a
correlation coefficient $r=0.69\pm{}0.12$ is obtained. To our knowledge, this
constitutes the first experimental determination of the correlation
coefficient between neighboring hydrogen bonds in a liquid, a
fundamental parameter that characterizes hydrogen-bond cooperativity
in a quantitative way. As yet, no theoretical prediction is available
for this number, and we hope the results presented here will stimulate
work in this direction.

\begin{acknowledgements}
  The author gratefully acknowledges Huib Bakker, Jennifer Herek and
  Wybren Jan Buma for critically reading the manuscript, and Wietze
  Buster for technical support.
\end{acknowledgements}


\begin{thebibliography}{28}
\bibitem[{met({\natexlab{a}})}]{meth_noot1}
\bibinfo{note}{We have also calculated the first-order correction when using
  anharmonic vibrational wavefunctions. Using a Morse potential with parameters
  determined from the $v=0,1,2$ levels (and the expressions given in
  Ref.~[\onlinecite{tipping83}]), we obtain
  $\langle{}11|q_1q_2|11\rangle=0.0097$. Hence, the more precise first-order
  correction $\langle{}11|\beta{}q_1q_2|11\rangle$ is not exactly zero, but
  still only 1\% of the second-order correction.}
\bibitem[{met({\natexlab{b}})}]{meth_noot2}
\bibinfo{note}{The information obtained from the simultaneous absorption
  spectrum is to some extent comparable to that derived from two-dimensional
  optical spectroscopy methods~\cite{mukamel00,stolow04}, which can also be
  used to determine couplings between
  vibrations~\cite{peter99,golonzka01,krum06} and even correlations between
  vibrational frequencies~\cite{demir02}. However, since two-dimensional
  vibrational spectroscopy requires that the coupled vibrations have different
  center frequencies, it cannot be used to reveal the molecular interaction and
  hydrogen-bond correlation observed here.}
\end{thebibliography}
\end{document}